\documentclass[sigconf]{acmart}

\AtBeginDocument{%
  }

\copyrightyear{2023}
\acmYear{2023}
\setcopyright{acmlicensed}
\acmConference[MM '23] {Proceedings of the 31st ACM International Conference on Multimedia}{October 29--November 3, 2023}{Ottawa, ON, Canada.}
\acmBooktitle{Proceedings of the 31st ACM International Conference on Multimedia (MM '23), October 29--November 3, 2023, Ottawa, ON, Canada}
\acmPrice{15.00}
\acmISBN{979-8-4007-0108-5/23/10}
\acmDOI{10.1145/3581783.3612187}

\acmSubmissionID{2176}

\usepackage{stfloats}
\usepackage[ruled,linesnumbered]{algorithm2e}

\usepackage{bbm}

\begin{document}

\title{Dynamic Low-Rank Instance Adaptation for Universal Neural Image Compression}


\author{Yue Lv}
\email{lvy21@mails.tsinghua.edu.cn}
\affiliation{%
  \institution{Tsinghua Shenzhen International Graduate School, Tsinghua University}
  \city{Shenzhen}
  \country{China}}
  \authornote{Equal contribution.}  \authornote{Work done during an internship at Tencent AI Lab.}

\author{Jinxi Xiang}
\email{jinxixang@tencent.com}
\affiliation{%
  \institution{Tencent AI Lab}
  \city{Shenzhen}
  \country{China}}
\authornotemark[1]

\author{Jun Zhang}
\email{junejzhang@tencent.com}
\affiliation{%
  \institution{Tencent AI Lab}
  \city{Shenzhen}
  \country{China}}
  \authornote{Corresponding authors.}
  
\author{Wenming Yang}
\email{yang.wenming@sz.
tsinghua.edu.cn}
\affiliation{%
  \institution{Tsinghua Shenzhen International Graduate School, Tsinghua University}
  \city{Shenzhen}
  \country{China}}
\authornotemark[3]

\author{Xiao Han}
\email{haroldhan@tencent.com}
\affiliation{%
  \institution{Tencent AI Lab}
  \city{Shenzhen}
  \country{China}}

\author{Wei Yang}
\email{willyang@tencent.com}
\affiliation{%
  \institution{Tencent AI Lab}
  \city{Shenzhen}
  \country{China}}

\renewcommand{\shortauthors}{Yue Lv et al.}

\begin{abstract}
The latest advancements in neural image compression show great potential in surpassing the rate-distortion performance of conventional standard codecs. Nevertheless, there exists an indelible domain gap between the datasets utilized for training (i.e., natural images) and those utilized for inference (e.g., artistic images).
Our proposal involves a low-rank adaptation approach aimed at addressing the rate-distortion drop observed in out-of-domain datasets. Specifically, we perform low-rank matrix decomposition to update certain adaptation parameters of the client's decoder. These updated parameters, along with image latents, are encoded into a bitstream and transmitted to the decoder in practical scenarios. Due to the low-rank constraint imposed on the adaptation parameters, the resulting bit rate overhead is small.
Furthermore, the bit rate allocation of low-rank adaptation is \emph{non-trivial}, considering the diverse inputs require varying adaptation bitstreams. We thus introduce a dynamic gating network on top of the low-rank adaptation method, in order to decide  which decoder layer should employ adaptation. The dynamic adaptation network is optimized end-to-end using rate-distortion loss.
Our proposed method exhibits universality across diverse image datasets. Extensive results demonstrate that this paradigm significantly mitigates the domain gap, surpassing non-adaptive methods with an average BD-rate improvement of approximately $19\%$ across out-of-domain images. 
Furthermore, it outperforms the most advanced instance adaptive methods by roughly $5\%$ BD-rate.
Ablation studies confirm our method's ability to universally enhance various image compression architectures.
Our project is available at \url{https://github.com/llvy21/DUIC}.
\end{abstract}





\maketitle


\section{Introduction}
Image compression is a fundamental technology for media storage and transmission in the digital information era. Common compression standards are JPEG \cite{wallace1992jpeg}, JPEG2000 \cite{skodras2001jpeg}, BPG \cite{bpg}, and the latest VVC \cite{bross2021overview}. With the development of deep learning, neural image  codecs have led to several breakthroughs in this area. Recent studies are on par with, or even outperform the most advanced VVC codec in terms of rate-distortion performance \cite{cheng2020learned, qian2022entroformer, he2022elic, he2021checkerboard, wang2023evc}.

\begin{figure}[t]
\includegraphics[width=\linewidth]{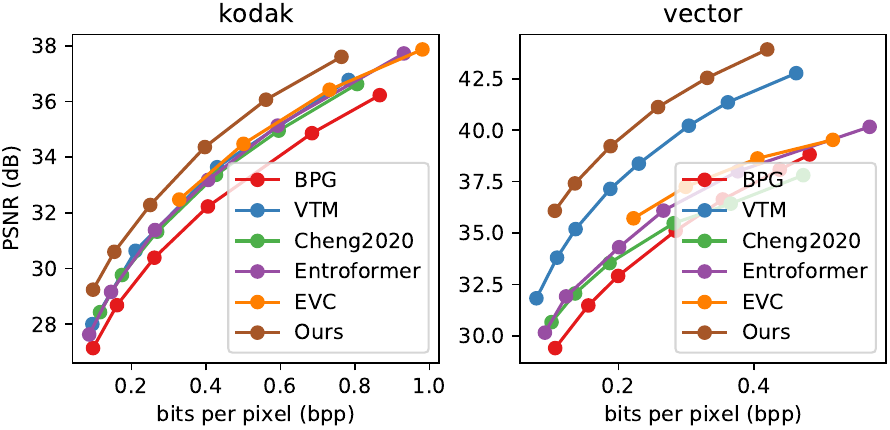}
\includegraphics[width=\linewidth]{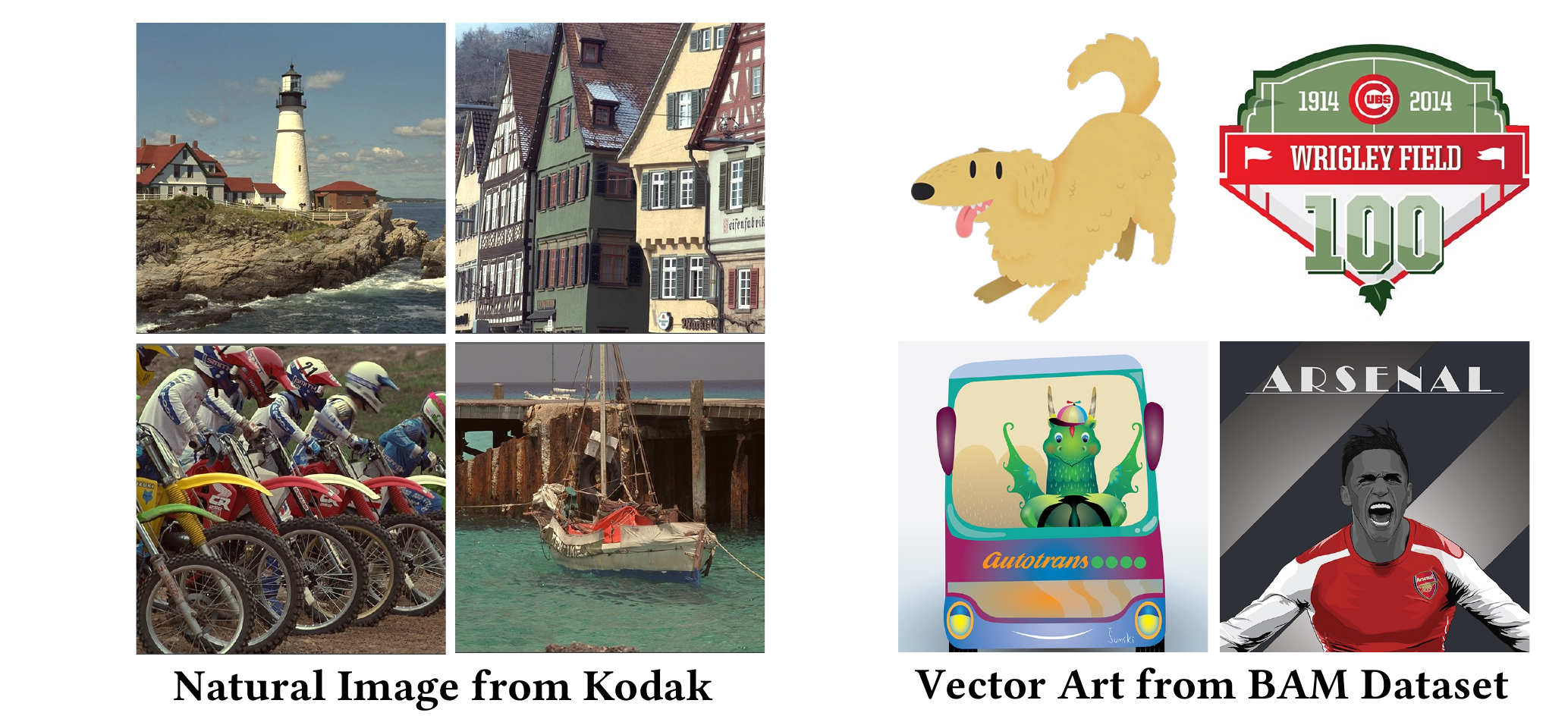}
\caption{(upper left) Three state-of-the-art neural image codecs from Cheng2020 \cite{cheng2020learned}, Entroformer \cite{qian2022entroformer}, and EVC \cite{wang2023evc}  perform fairly well on the Kodak dataset. (upper right) But their performance drops significantly on an out-of-domain dataset containing artistic images. We argue that neural codecs trained on natural images are imperfect from the perspective of generalization. \textit{We aim to solve this bottleneck with a dynamic adaptation design without updating all the parameters of the codec for each specialized domain. } Our proposed instance adaptation method based on Cheng2020 shows its robustness on out-of-domain images.
}
\label{fig:deterior}
\end{figure}

It is a common practice to train neural image codecs on natural images, such as CLIC \cite{clic}  and
OpenImage \cite{kuznetsova2020open}, and then evaluated on natural images benchmarks, such as Kodak \cite{kodak}, Tecnick \cite{asuni2014testimages}, and DIV2K \cite{agustsson2017ntire}.  Apart from natural images, there exists a diverse, relatively unexplored space of artistic drawing, comics, and gaming which offer depictions of the world through artwork and are transmitted through the internet every day. Extending neural compression models trained with neural images to these domains is non-trivial. To highlight this issue, we showcase three representative models from Cheng2020 \cite{cheng2020learned}, Entroformer \cite{qian2022entroformer}, and EVC \cite{wang2023evc} in Fig. \ref{fig:deterior}.  The results show that neural image codecs are comparable with VVC on the Kodak dataset, and consistently outperform the BPG. However,  the RD performance degrades dramatically on an artistic dataset from BAM \cite{wilber2017bam}, legging behind the BPG in general. Although neural codecs have proven to be very successful, a model trained to minimize expected RD cost over natural images is unlikely to be optimal for both optimization and generalization. The problem of generalization would be critical when the testing distribution differs from the training distribution.

\begin{figure}[t]
\centering
\includegraphics[width=0.9\linewidth]{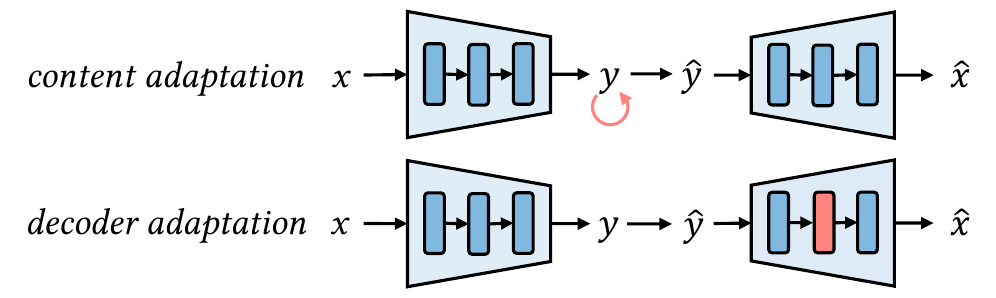}
\caption{Different approaches for compression adaptation.}
\label{fig:instance_adapt}
\end{figure}

To enhance the compression quality of out-of-domain images during inference, instance-adaptive compression methods have been suggested, as shown in Fig. \ref{fig:instance_adapt}. 
One line of work focuses on \textit{content adaptation}, which adjusts the neural codecs at inference time by refining the latent representation extracted from the encoder. The values of the latent of the image are directly changed during the encoder transformation stage, making the modified latent representation fit the decoder \cite{campos2019content, guo2020variable, yang2020improving}. 
This method is attractive because it does not require any extra information to be encoded in the bitstream, and no alterations will be made to the receiver's end. However, the benefits in terms of performance are restricted since neither the encoder nor the decoder can be adjusted.

Another approach, which is complementary to latent adaptation, is \textit{decoder adaptation} by fine-tuning specific parameters of the decoder. 
The scenario is that the decoder is frozen when deployed at the client's end, but we can adjust certain parameters by transmitting specific bitstreams.
Fine-tuning more network parameters brings better performance for sure, but it also requires more rate cost.
Rozendaal \textit{et al.}\cite{van2021overfitting, van2021instance} fine-tuned all the parameters and compress the model updates in spike-and-slab prior. Although their approach achieves some degree of dynamic fine-tuning of the number of parameters, the way of updating parameters lacks efficiency and fails to apply to images for a huge extra bit cost. 
Existing methods \cite{lam2020efficient, van2021instance, van2021overfitting, zou2021adaptation} overlook the importance of designing a \textbf{parameter-efficient} approach  to maximize compression performance while minimizing extra bitrate.
Essentially, fine-tuning parameters should also be treated as a rate-distortion problem. 
Tsubota \textit{et al.}\cite{tsubota2023universal} was the first to propose updating the model in a parameter-efficient approach and introduced 768 extra parameters. 
Although extra parameters were optimized with rate-distortion loss, the number of updated parameters cannot be tailored to the content. This implies that the rate is fixed without flexibility.

 


We propose a universal adaptation method for neural image compression to efficiently update the decoder parameters during inference. 
Our approach involves utilizing a plug-and-play adaptation block, which is composed of two low-rank matrices, in order to minimize the additional bit costs associated with updating the model. To optimize the adaptive parameters of an instance up to its theoretical upper limit, we have developed a flexible and dynamic decoder gate network that can adapt based on the number and position of blocks to be plugged in. This approach enables the model to automatically determine the bit rate budget of each layer, resulting in significant rate-distortion gains.

Our contributions can be summarized as follows:
\begin{itemize}
    \item We propose to update the decoder weights using additional low-rank decomposition matrices at inference time, which are compressed and transmitted along with the image. 
    \item      Our proposed dynamic gating network has the ability to learn the optimal number and positions of adaptation automatically, facilitating efficient domain transfer. 
    \item  This technique effectively minimizes the domain gap between training and inference, while also providing a versatile approach toward instance-adaptive compression. Empirical evidence indicates that our approach can be extended to various other neural image compression models.
\end{itemize}

\begin{figure*}
\centering
\includegraphics[width=175mm]{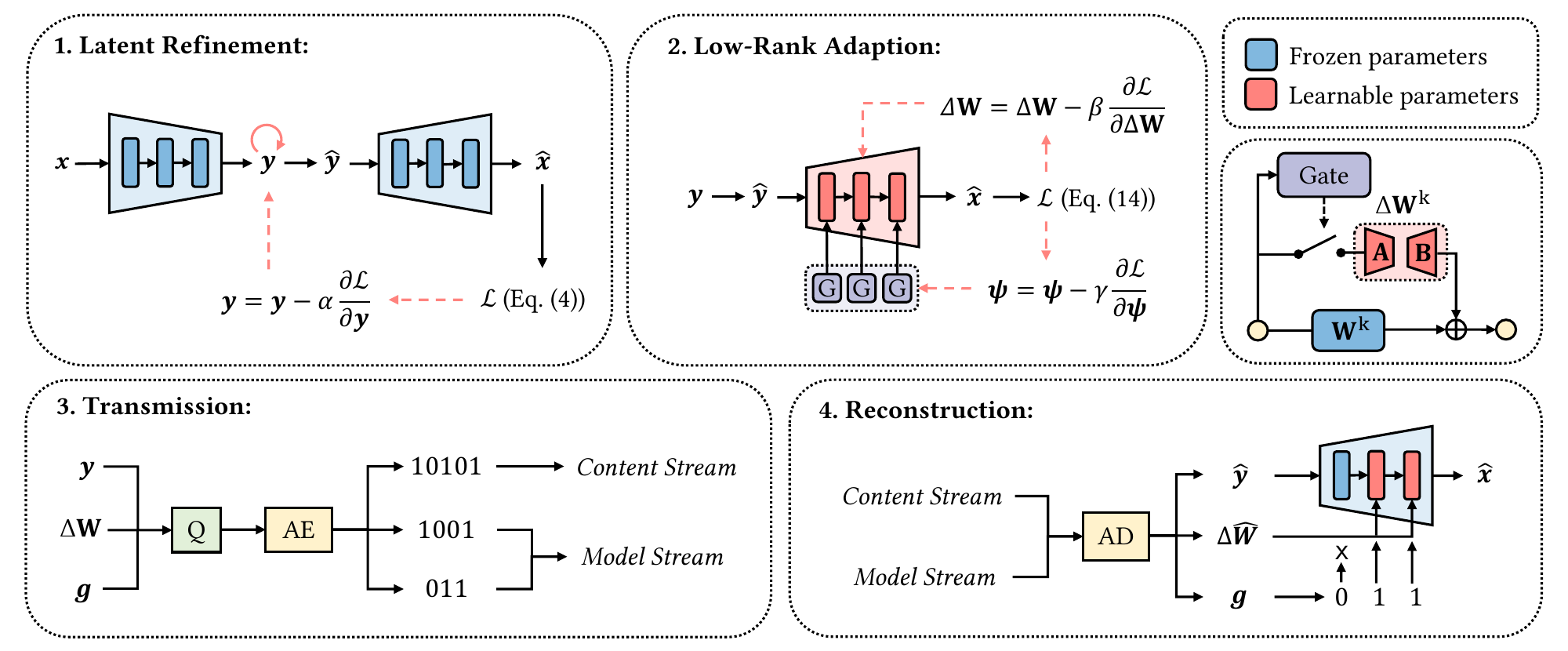}
\caption{Simplified pipeline of the dynamic universal instance adaptive image compression. Steps 1-3 are executed on the server and step 4 is executed on the client. 
1) We first refine the latent representation $\boldsymbol{y}$ with all the parameters fixed. 
2) Then we adapt the model by dynamically choosing the layers to update.
Incremental model update parameters $\Delta \mathbf{W}$ and the gate networks $\boldsymbol{G_{\psi}}$ are optimized in this step.
3) The latent code $\boldsymbol{y}$, instance incremental matrix $\Delta \mathbf{W}$, and the gate output $\boldsymbol{g}$ are quantized and encoded as content stream and model stream, respectively. 4) Finally, the client decodes the bit stream and updates the decoder according to the gate binary decisions to reconstruct the latent code.}
\label{fig:pipeline}
\end{figure*}

\section{Related Work}

\subsection{Neural image compression}
Neural compression is based on an autoencoder that tries to reconstruct the input instance from quantized latents, cooperated with a prior that is used to losslessly compress these latents. The transform coding paradigm can be formulated by:
\begin{align}
\boldsymbol y  &=g_a(\boldsymbol x; \boldsymbol{\phi}) \\ 
\boldsymbol{\hat y} &=Q(\boldsymbol{y}) \\ 
\boldsymbol{\hat x} &=g_s(\boldsymbol{\hat y}; \boldsymbol{\theta} ) \label{eq:gs} 
\end{align}
where $\boldsymbol x$ is input original image; $\boldsymbol y$ is latent code generated by analysis transform $g_a$ with the optimized parameter $\phi$;  $\boldsymbol{\hat y}$ is quantized latent code;  $\boldsymbol{\hat x}$ is compress reconstructed images by synthesis transform $g_s$ with the parameter $\theta$.
A lossy neural image compression network is optimized with rate-distortion loss:
\begin{equation}
     \mathcal L = \mathcal R + \lambda \cdot \mathcal  D = -\log p(\boldsymbol{\hat y}) 
     + \lambda \cdot \mathcal  D (
     \boldsymbol{\hat x}, \boldsymbol{x}), \label{eq:loss}
\end{equation}
where $\mathcal{R}$ is the bit rate; $\mathcal{D}$ measures the image reconstruction distortion; $\lambda$ is the Lagrange multiplier controlling R-D trade-off. 

To improve the compression performance, recent studies focus on exploring more powerful non-linear neural network modules for image transformation, such as residual networks \cite{cheng2020learned,he2022elic}, self-attention \cite{chen2021end}, ConvNext \cite{duan2023lossy}, and transformer \cite{lu2021transformer,zhu2022transformer,zou2022devil,xiangmimt}. 

Another research direction is to develop accurate entropy models to reduce redundancy, such as hierarchical \cite{balle2018variational}, unidirectional auto-regressive model \cite{minnen2018joint, minnen2020channel}, checkerboard auto-regressive \cite{he2021checkerboard, he2022elic}, and bi-directional auto-regressive \cite{xiangmimt}.
Hierarchical prior employs no context information and the performance is limited. Unidirectional auto-regressive is a context model that predicts the next code with all the decoded neighbor pixels serially and achieves superior performance. To accelerate the auto-regressive, channel-wise \cite{minnen2020channel}, spatial-wise \cite{he2021checkerboard}, spatial and channel \cite{he2022elic}, and random decode sequence \cite{xiangmimt} were proposed to prompt the parallel coding. Some other works introduce codebooks and vector-quantize the latent code to compress\cite{zhu2022unified} or use hierarchical VAE to compress latent code\cite{duan2023lossy}. 

Although the above-mentioned designs push the boundaries of compression further, the learning-based components are not fully investigated on out-of-domain datasets to verify their robustness.

\subsection{Instance Adaptive Image Compression}



To address the issue of the domain gap, several notable studies have been conducted. One approach involves the fine-tuning of the encoder during inference. This method aims to correct the latent code without introducing additional bit transmission. For example, Campos \textit{et al.} \cite{campos2019content}  solved it through an iterative procedure similar to during training with gradient descent applied on latents. Guo \textit{et al.} \cite{guo2020variable} further added side information after the latent representation has been refined.
Yang \textit{et al.} \cite{yang2020improving}  extended bits-back coding by encoding  some side information using an entropy model.

On the other hand, certain studies have focused on fine-tuning the decoder during inference.
For example, Lam \textit{et al.}\cite{lam2020efficient} update the deviation parameters of the convolutional layer in the decoder and compress the deviation parameters with the image for transmission; Zou \textit{et al.}\cite{zou2021adaptation} use a set of overfittable multiplicative parameters (OMP) to overfit a video. The set of OMP is transmitted together with the images; 
Liu \textit{et al.} \cite{liu2021overfitting} downsample the video and then overfit a video super-resolution network and transmit the parameters of the super-resolution network together with the video; 
Tsubota \textit{et al.} \cite{tsubota2023universal} propose to update the model in a parameter-efficient approach using an adapter. 
Besides, a few studies fine-tune the whole model. Rozendaal \textit{et al.} \cite{van2021overfitting,van2021instance} overfit the whole model for the input data, using spike-and-slab probability distribution to compress the entire decoder and entropy model since the client isn't affected by the encoder update. 

Instance-adaptive image compression is connected with parameter-efficient transfer learning. To optimize compression performance and minimize bitstream overhead, we should adopt a parameter-efficient approach due to the need to compress and transmit model update parameters as extra bits.
Parameter efficient transfer learning origins from the  pre-trained large language models \cite{devlin2018bert, radford2019language}.
Studies were proposed to adapt a large-scale pre-trained model to downstream tasks by updating only a small number of parameters. For example, adapter tuning \cite{houlsby2019parameter} inserts  bottleneck structure between existing layers; prefix tuning \cite{li2021prefix} and prompt tuning \cite{lester2021power} attach additional prefix tokens to the input; diff pruning \cite{guo2020parameter} incremental update the pre-trained weights as a sparse matrix. LoRA \cite{hu2021lora} utilizes the low intrinsic dimension in the model and parameterizes incremental weight as a low-rank matrix by the product of two small matrices achieves comparable or even better performance than full fine-tuning.

\section{Method}

\subsection{Overview}
The proposed method is illustrated in Fig. \ref{fig:pipeline}.
With a pre-trained neural image compression model, we freeze the base model and incorporate additional image-specific parameters to the decoder. 


To begin, the encoder extracts the latent code from the input image, which is then refined with rate-distortion loss following a simplified approach of Yang \textit{et al.} \cite{yang2020improving}. The latent code $\boldsymbol{y}$ is optimized by gradient descent with frozen all the pre-trained weights. After latent refinement, the weight of the decoder is updated in an adaptive, parameter-efficient approach. Both the latent code and model update are encoded and transmitted from the server to clients in the content stream and model stream.

During the decompression process, the client decodes both the content and model streams and subsequently updates a small portion of the decoder with instance-specific parameters. This enables the reconstruction of image latent code using the updated decoder. 

\subsection{Low-Rank Adaptation}

For a pre-trained weight matrix of a specific layer $\mathbf{W}^k \in \mathbb{R}  ^{c_{out} \times c_{in} \times h \times w}$, we aim to adapt it for universal image compression by modifying its weight in an efficient way:
\begin{equation}
    \boldsymbol{h}^{k+1}  = (\mathbf{W}^k  + \Delta\mathbf{W}^k )\boldsymbol{h}^{k}
    \label{eq:param_update}
\end{equation}
where $\boldsymbol{h}^{k}, \boldsymbol{h}^{k+1}$ are  the $k$-th, and $(k+1)$-th layer of features.
We  leverage the low-rank property of $\Delta\mathbf{W}^k$ to update instance-specific parameters in an efficient manner, ensuring maximum reconstruction quality with minimal bit overhead. 
We freeze $\mathbf{W}^k$ and incrementally update it with two learnable matrices $\mathbf{A}$ and $\mathbf{B}$ by low-rank decomposition:
\begin{equation}
     \Delta \mathbf{W}^k = \mathbf{B} \mathbf{A}
\end{equation}
where $\mathbf{A} \in \mathbb{R}^{r \times c_{in}}$ is initialized with random Gaussian, $\mathbf{B} \in \mathbb{R}^{c_{out}\times r}$ is initialized as zero. This initialization ensures $\Delta \mathbf{W}^k=0$ at the beginning of adaptation. The rank $r \ll \min \{c_{in}, c_{out}\}$. 


We note ${\Delta \boldsymbol{\theta}} = \{\mathbf{A}, \mathbf{B}\}$. During the model update, $\Delta \boldsymbol{\theta}$ is quantized with quantization interval $w$.
As there is no prior for $\Delta \boldsymbol{\theta}$, we use a factorized density model $\boldsymbol{\pi}$ to encode it into bitstream:
\begin{equation}
    p_{\Delta \hat{\boldsymbol{\theta}} \mid \boldsymbol{\pi}} (\Delta \hat{\boldsymbol{\theta}} \mid \boldsymbol{\pi})=\prod_i\left(p_{\Delta \theta_i \mid \boldsymbol{\pi}}(\boldsymbol{\pi}) * \mathcal{U}\left(-\frac{w}{2}, \frac{w}{2}\right)\right)\left(\Delta  \hat{\theta}_i\right)
    \label{eq:factorized_w}
\end{equation}
where $\Delta \theta_i$ denotes the $i$-th element of $\Delta \boldsymbol{\theta}$ with $i$ as the position of each element; $\mathcal{U}\left(-\frac{w}{2}, \frac{w}{2}\right)$ is a uniform noise as a stand-in for quantization. We specify $\boldsymbol{\pi}$ as a logistic distribution. As  such, the probability estimation for $\Delta \hat{\boldsymbol{\theta}}$ is:
\begin{equation}
    p_{\boldsymbol{\pi}}\left(\Delta \hat{\theta}_i \right) =c\left(\Delta \hat{\theta}_i+\frac{w}{2}\right)-c\left(\Delta \hat{\theta}_i-\frac{w}{2}\right)
\label{eq:cdf_theta}
\end{equation}
with $c()$ as the cumulative distribution function of $\boldsymbol{\pi}$
calculated by logistic distribution:
\begin{equation}
c(d) = \frac{1}{2}+\frac{1}{2}\tanh{\left(\frac{d-\mu}{2s}\right)}
\end{equation}
where $w$, $\mu$, $s$ are predefined parameters. We set $w=0.01$, $\mu=0$, $s=0.05$ by default.

Combining with the rate of latents $\mathcal R(\hat {\boldsymbol{ {y}}})$ and distortion of reconstructed images $\mathcal D(\boldsymbol{\hat x},\boldsymbol{x})$, the low-rank adaptation is optimized end-to-end with:
\begin{align}
    \begin{aligned}
    \mathcal L & =\mathcal R (\hat {\boldsymbol{ {y}}}) +\mathcal{R}(\Delta {\hat{\boldsymbol{\theta}}}) + \lambda \cdot \mathcal D(\boldsymbol{\hat x},\boldsymbol{x}\mid \boldsymbol{\theta}, \Delta\hat{\boldsymbol{\theta}} ) \\
    & = 
     - \log p(\hat {\boldsymbol{ {y}}})
     - \log p_{\boldsymbol{\pi}}(\Delta \hat{\boldsymbol{{\theta}}}) 
     + \lambda \cdot \mathcal D(g_s(\hat {\boldsymbol{ {y}}} ; \boldsymbol{\theta}, \Delta \hat {\boldsymbol{\theta}}),\boldsymbol{x}) 
    \end{aligned}
     \label{eq:rrd}
\end{align}
with $\lambda$ as the multiplier for the trade-off between rate and distortion.

The low-rank adaptation of Eq. \eqref{eq:param_update} can be readily applied to various decoder layers, such as convolutional layers and attention layers. In typical compression models \cite{cheng2020learned, he2022elic}, one decoder layer, such as a $3 \times 3$ convolutional layer, contains a large number of parameters, specifically $C \times C \times 3 \times 3$, where $C$ denotes the number of channels. By utilizing the low-rank adaptation technique, the number of learnable parameters can be significantly reduced to $C \times r \times 2$, resulting in substantial savings in terms of extra bitrate.

\label{sec:lora}

\subsection{Dynamic Gating Network}

\begin{figure}
\centering
\includegraphics[width=0.75\linewidth]{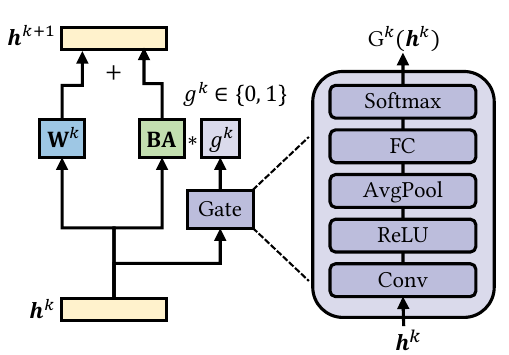}
\caption{Dynamic gate network structure.}
\label{fig2:env}
\end{figure}

Eq. \eqref{eq:rrd} delineates the rate-distortion loss for optimizing a single layer in low-rank adaptation. It is a logical progression to expand the use of low-rank adaptation across multiple layers:
\begin{equation}
    \mathcal{L} = \mathcal{R}(\hat{\boldsymbol{y}}) + \sum_{k=1}^{K}\mathcal{R}(\Delta \hat{\boldsymbol{\theta}}_k) + \lambda\cdot \mathcal{D}(\hat{\boldsymbol{x}}, \boldsymbol{x}\mid \boldsymbol{\theta},  \Delta \hat{ \boldsymbol{\theta}}_1,\ldots,\Delta\hat{\boldsymbol{\theta}}_K  )
    \label{eq:multiple_layer}
\end{equation}
where $K$ is the total number of layers with low-rank adaptation in the decoder. Quantifying the number of layers utilizing low-rank adaptation presents a challenge due to (1) the varying adaptation levels required for each instance and (2) the differing rate-distortion enhancements with diverse bitstream allocation strategies.

To augment the flexibility and adaptability of our instance adaptive image compression model, we devise a dynamic architecture that enables adaptive control over the number and location of plug-in blocks inspired by \cite{wang2018skipnet}. In contrast to a static network that employs the same architecture during inference, our proposed dynamic instance adaptive network can dynamically adjust its own structure to suit various raw inputs during processing.

We explicitly assign a gating network $G$ to each incremental matrix. The gating module maps the previous layer outputs to the binary decision to skip or execute the weight update. We note the number of the insertable positions for an incremental update in synthesis transform as $k \in \{1, 2, 3, \cdots, K\}$, for the $k$-th input $\boldsymbol{h}^k$, the incremental update is controlled by the gating signal $g^k \in \{0,1\} $, then we define the output of the gated incremental layer as 
\begin{equation}
    \boldsymbol{h}^{k+1}  = (\mathbf{W}^k + g^k \Delta \mathbf{W}^k)\boldsymbol{h}^k
\end{equation}


During the forward pass, we apply hard-gating to the gate outputs, which are relaxed to soft-gating during backpropagation:
\begin{equation}
    g^k = \text{sg}\left[ \mathbbm{1}( G^k (\boldsymbol{h}^k) \geq 0.5 )\right] + G^k(\boldsymbol{h}^k) - \text{sg}\left[ G^k(\boldsymbol{h}^k) \right]
\end{equation}
where $\mathbbm{1}$ is the indicator function; $G^k$ is the gating network; \text{sg} stands for the stop gradient operator that is defined as an identity at forward computation time
and has zero partial derivatives.

We implement a dynamic adaptive gate network (shown in Fig. \ref{fig2:env}) that comprises a convolution layer, ReLU activation, adaptive average pooling, full connection, and softmax layer. The gating network maps the input to a binary output $\{0, 1\}$ that dynamically controls the layers of adaptation.


We optimize Eq. \eqref{eq:multiple_layer}  with the dynamic gating network as:
\begin{equation}
    \begin{aligned}
\mathcal{L} = & \mathcal{R}(\hat{\boldsymbol{y}}) + \sum_{k=1}^{K} g^k\cdot \mathcal{R}( \Delta \hat{\boldsymbol{\theta}}_k)  \\
& + \lambda\cdot \mathcal{D}(\hat{\boldsymbol{x}}, \boldsymbol{x}\mid \boldsymbol{\theta},  g^1\Delta  \hat{ \boldsymbol{\theta}}_1,\ldots,g^K\Delta\hat{\boldsymbol{\theta}}_K  )
    \end{aligned}
\end{equation}


The algorithm for dynamic low-rank adaptation in image compression is presented in Algorithm \ref{alg:1}. The latent representation $\boldsymbol{y}$ is initially refined for $N_1$ steps, following established practices \cite{yang2020improving, tsubota2023universal}. 
Subsequently, the decoder parameters are adapted with low-rank constraints for additional $N_2$ steps. It is recommended to warm up each low-rank layer $\Delta \boldsymbol{\theta}_k$ for the $N_w$ steps to prevent deterioration in reconstruction performance during the early stages.

\begin{algorithm}[t]
\caption{Dynamic Low-Rank Adaptation for Image Compression}\label{alg:1}
\KwIn{input image $\boldsymbol{x}$, latent refinement step $N_1$, model adaptation step $N_2$, 
warm-up step $N_w$.}
\KwOut{reconstructed image $\boldsymbol{\hat x}$, model updates $\{ G^1, \Delta \boldsymbol{\theta}_1, \ldots G^K, \boldsymbol{\theta}_K\} $.}
$n \gets 0$\;
$\boldsymbol y  \gets g_a(\boldsymbol{x} ; \boldsymbol{\phi})$\;
\tcc{Content adaptation, refine the latent
$\boldsymbol{y}$} 
\While{$n < N_1$}{ 
$\boldsymbol{\hat y} \gets Quant(\boldsymbol{y})$\;
$\mathcal{L}(\boldsymbol{y}) \gets -\log p(\boldsymbol{\hat y}) + \lambda \cdot \mathcal{D}(\boldsymbol{x}, g_s(\boldsymbol{\hat y}; \boldsymbol{\theta}))$\;
$\boldsymbol{y} \gets \boldsymbol{y} - \alpha \frac{\partial \mathcal{L}}{\partial \boldsymbol{y}}$\;
$n \gets n+1$\; 
}
$n \gets 0$\;
$\boldsymbol{\hat y} \gets Quant(\boldsymbol{y})$\;
\tcc{Decoder adaptation ${G^1,\ldots
G^K }$ and $\boldsymbol{\Delta \theta}^1, \ldots \boldsymbol{\Delta \theta}^K$} 
\While{$n < N_2$}{ 
    $\boldsymbol{g} \gets \boldsymbol{1} $\textbf{\ if\ }$ n \leq N_w$ \textbf{else} $ G_\psi(g_s,\boldsymbol{\hat y};\boldsymbol{\theta}, \boldsymbol{\Delta \theta})$; \tcp{warm up}
    $\mathcal{L} \gets -\log p(\boldsymbol{\hat y}) -\log p_{\boldsymbol{\pi}}(\boldsymbol{\Delta \hat \theta}) + \lambda \cdot \mathcal{D}(\boldsymbol{x}, g_s(\boldsymbol{\hat y}; \boldsymbol{\theta}, \boldsymbol{g} \boldsymbol{\Delta \hat \theta }))$\;
    $\boldsymbol{\Delta \theta } \gets \boldsymbol{\Delta \theta } - \beta \frac{\partial \mathcal{L}}{\partial \boldsymbol{\Delta \theta}}$; \tcp{optimize model updates}
    $\psi \gets \psi - \gamma \frac{\partial \mathcal{L}}{\partial \psi} $;  \tcp{optimize gate network}
    $n \gets n+1$\; 
}

 \end{algorithm}
\section{EXPERIMENTS}

\subsection{Experimental Setup}

{\bfseries Datasets.} 
Our method was evaluated on universal image benchmark datasets constructed in \cite{tsubota2023universal, wilber2017bam}. The dataset includes a diverse range of image types, including (i) 24 natural images from the Kodak dataset; (ii) 100 comics images (BAM); (iii) 100 vector art images (BAM); (iv) 100 pixel-style gaming images (self-collected). It is noteworthy that all image codecs were pre-trained on natural images. Therefore, we consider natural images as in-domain evaluations and other images as out-of-domain evaluations.


{\bfseries Training Recipe.} To assess the suggested adaptation approach, we select several pre-trained image codecs, namely Cheng2020 \cite{cheng2020learned}, Entroformer \cite{qian2022entroformer}, and EVC \cite{wang2023evc}. We prespecify $r=2$. The latent representation is first refined for $N_1 = 2,000$ steps, followed by an additional $N_2 = 2,000$ steps of optimization of the decoder's updated parameter. 
The learning rate for latent refinement and model update is set at $10^{-3}$, while for the gate network, it is $10^{-5}$.



\subsection{Rate-Distortion Performance}

\begin{figure*}
\centering
\includegraphics[width=0.4\linewidth]{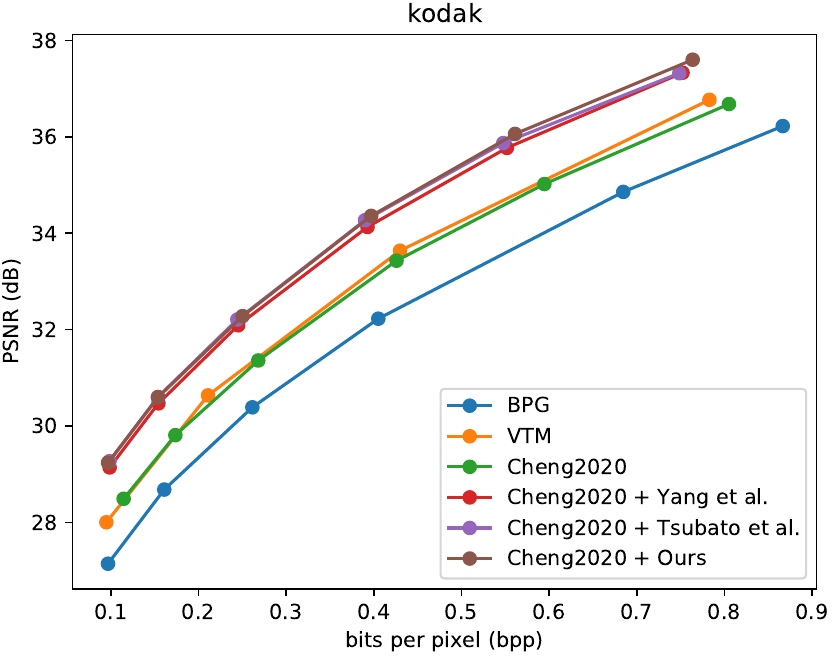}
\ \ \ \ \ \ \ \ \ \ \ \ 
\includegraphics[width=0.4\linewidth]{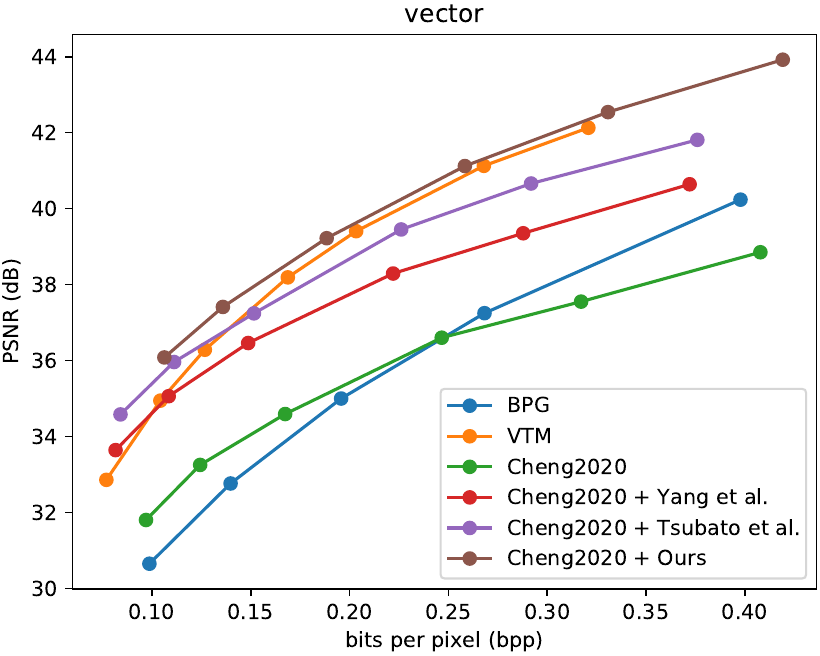}
\\ 
\includegraphics[width=0.4\linewidth]{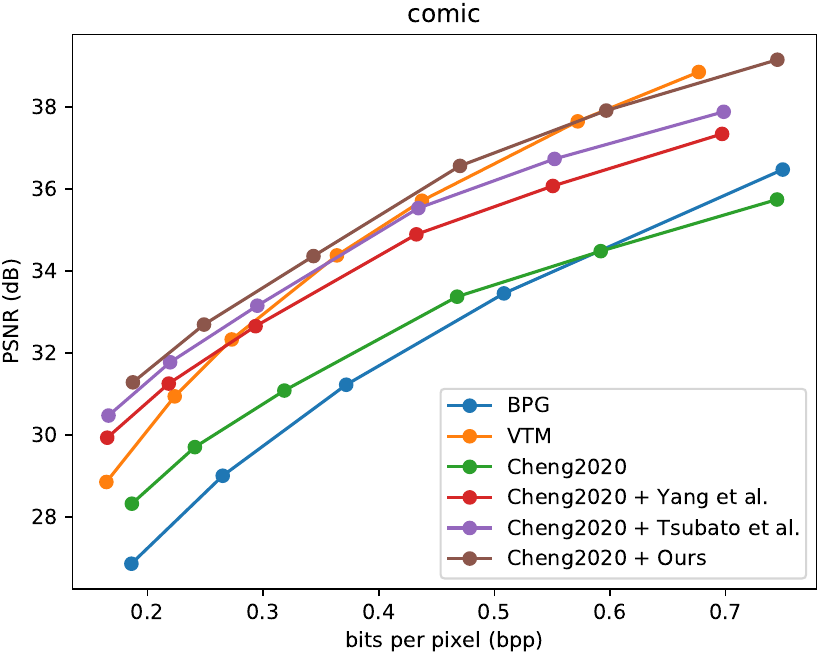}
\ \ \ \ \ \ \ \ \ \ \ \ 
\includegraphics[width=0.4\linewidth]{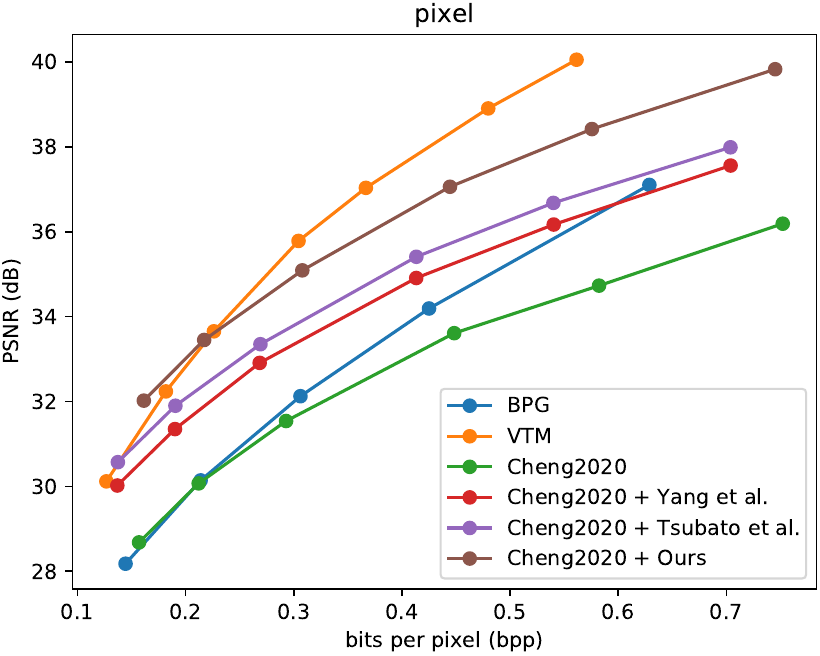}
\caption{The RD performance with BPG, VVC, and existing instance-adaptive image compression methods Tsubota \textit{et al.} \cite{tsubota2023universal} and Yang \textit{et al.} \cite{yang2020improving} based on the same pre-trained codec \cite{cheng2020learned}. BD rates are shown in Table \ref{tab:overall}.}
\label{fig:overall}
\end{figure*}

\begin{table*}
  \caption{Our method is compared with existing adaptive approaches, as well as BPG and VVC. The evaluation is measured in terms of BD-rate (\%) using VVC as the anchor. Smaller values indicate superior performance.}
  \label{tab:overall}
  \begin{tabular}{cccccc}
    \toprule
    Method&Nature image&Vector art&Comic& Pixel art&Average \\
    \midrule
    BPG                     
    & 9.57 & 18.92 & 11.98 & 18.29 & 14.69 \\
    VVC                     
    & 0.00 & 0.00 & 0.00 & $\boldsymbol{0.00}$ & 0.00 \\
    Cheng2020 \cite{cheng2020learned}            
    & 1.61 & 14.68 & 11.39 & 25.04 & 13.18 \\
    Cheng2020 + Yang \textit{et al.} \cite{yang2020improving}            
    & -6.90 & 3.32 & 4.06 & 12.75 & 3.31 \\
    Cheng2020 + Tsubota \textit{et al.}  \cite{tsubota2023universal}        
    &  -7.94 & 0.38 & 1.55 & 10.55 & 1.14 \\
    Cheng2020 + Ours                    
    & -$\boldsymbol{8.08}$ &  -$\boldsymbol{2.01}$ & -$\boldsymbol{1.13}$ & 6.36 & -$\boldsymbol{1.21}$\\
  \bottomrule
\end{tabular}
\end{table*}

We compare our method with open-source instance-adaptive image compression methods \cite{yang2020improving, tsubota2023universal} using the same pre-trained image codec cheng2020 \cite{cheng2020learned}. 
In our comparison, we also include traditional codecs such as BPG and VVC, as well as a baseline method without adaptation. To evaluate their performance, we calculate the peak signal-to-noise ratio (PSNR) and measure the data rate in bits per pixel (bpp). Subsequently, we plot the rate-distortion (RD) curves.
The experiment results are shown in Fig. \ref{fig:overall}. 
We observe that for Kodak images, all three  adaptation methods achieved comparable performance, resulting in approximately 1dB improvement in PSNR without domain gaps in natural images. 
For out-of-domain images such as vector art, comic, and pixel-style images, our method outperforms other adaptation methods and compensated for performance degradation. 
Notably, for out-of-domain comics and vector arts, our method even surpasses VVC. Nevertheless, the domain gap between pixel-style images and natural images proved too wide to fully compensate. Consequently, our method still significantly lags behind VVC, although it clearly outperforms other instance adaptive compression methods. 
Table \ref{tab:overall} presents a comparison of the Bjøntegaard delta (BD) bit-rate. In order to compute the BD rate, we use the VVC-intra as the anchor. Our method demonstrates superior performance in both in-domain and out-of-domain images when compared to other instance-adaptive compression methods. 

\subsection{Ablation Studies}

{\bfseries Low-Rank Adaptation v.s. Other Adaptations.} 
We investigate the decoder model update of the same convolution layer using the same pre-trained image codec \cite{cheng2020learned}.
We conduct a comparative study of our approach with other adaptation methods, namely bias-only \cite{lam2020efficient}, adapter \cite{tsubota2023universal}, fully fine-tuning \cite{van2021overfitting}, and Singular Value Decomposition (SVD) fine-tuning \cite{han2023svdiff} on out-of-domain images.
We update the $3\times 3$ conv layer in the decoder's penultimate layer using pre-trained weight matrix $\mathbf{W}^k \in \mathbb{R}  ^{c_{out} \times c_{in} \times h \times w}$, where $c_{out} = c_{in} = 128$ or $192$, $h = w = 3$ (i.e., kernel size). 
We compress the bias, singular value, adapter, and rank-composition matrices using a logistic distribution, while the fine-tuned incremental parameters are compressed using a spike-and-slab prior. 
We compare various model update approaches and their corresponding mathematical formulas, the number of parameters, and BD-rate. 
The number of updated parameters in the penultimate layer is calculated based on  Cheng2020 \cite{cheng2020learned} from compressai \cite{begaint2020compressai} at quality $q=6$.

\begin{figure}
\centering
\includegraphics[width=0.8\linewidth]{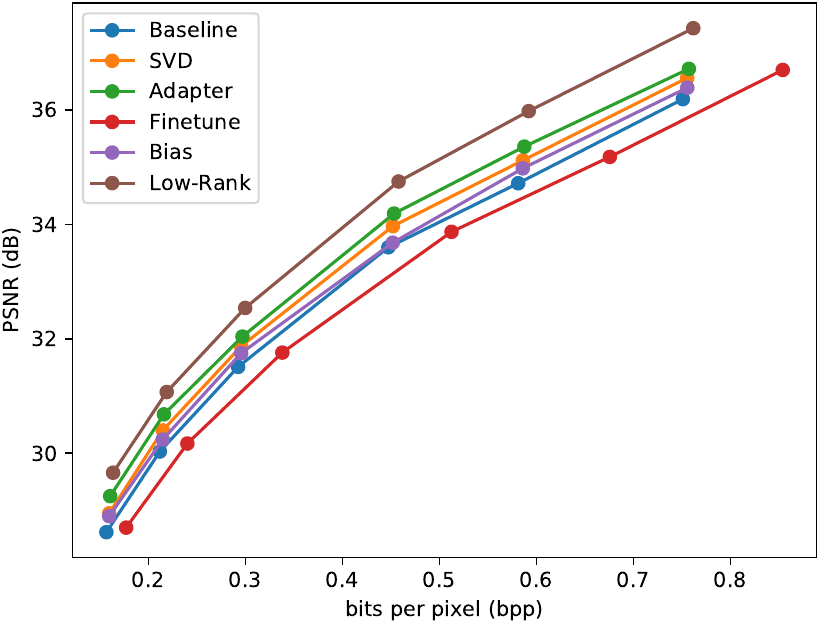} 
\caption{Comparison of model update approaches including SVD \cite{han2023svdiff}, adapter \cite{tsubota2023universal}, fine-tune all \cite{van2021overfitting}, bias-only \cite{lam2020efficient} and our low-rank update on pixel-style images.}
\label{fig:update}
\end{figure}

The experimental results are displayed in Fig. \ref{fig:update} and Table \ref{tab:update}. The findings indicate that parameter-efficient adaptations, such as SVD, adapter, and low-rank, demonstrate significantly better RD performance. Among these methods, updating the weight with low-rank decomposition yields the most significant improvement. In contrast, the bias-only method yields limited improvement, as it only applies to scenarios with minor disturbances. Fine-tuning all parameters in a convolution layer is not feasible due to high rate overhead and accompanying quantification errors. Additionally, updating the singular value of the pre-trained weight matrix provides limited enhancement with fixed orthonormal bases.

The adapter and low-rank weight updating techniques share some similarities. Both involve introducing a small number of learnable parameters, with the number of parameters being the same. They both use the matrix $\mathbf{A}$ to map the input representation to a low-dimensional space and then the matrix $\mathbf{B}$ to map back to the high-dimensional space. However, the adapter is executed serially on the hidden layer outputs, perturbing the outputs, while the low-rank update approach perturbs the pre-trained weights, resulting in more intense linear changes and stronger instance adaptation capabilities. Additionally, the adapter typically has larger values for its extra parameters, as they act on the output of one layer. In summary, the low-rank model update technique outperforms the adapter with equivalent update parameters.

\begin{table*}
  \caption{Compare different model update approaches within the same convolution layer with BD-rate (\%) and the number of updated parameters on vector art, comic, and pixel datasets.}
  \label{tab:update}
  \begin{tabular}{cccccccc}
    \toprule
    Update Approaches & Formulas & Updated params & $\#$ of params & Vector  & Comic & Pixel & Average \\
    \midrule
    baseline & $\boldsymbol{h} \leftarrow \mathbf{W}\boldsymbol{h} + \mathbf{b}$ & /  & 0 & 
    0.00 & 0.00 & 0.00 & 0.00 \\
    fine-tune  & $\boldsymbol{h} \leftarrow (\mathbf{W}+\Delta \mathbf{W})\boldsymbol{h} + (\mathbf{b} + \Delta \mathbf{b})$ & $\Delta \mathbf{W}$, $\Delta \mathbf{b}$ &   332k & 
    2.76 & 3.31 & 3.18 & 3.08 \\
    bias-only  & $\boldsymbol{h} \leftarrow \mathbf{W}\boldsymbol{h}+ (\mathbf{b}+\Delta \mathbf{b})$&  $\Delta \mathbf{b}$ & 192 & 
    -0.05 & -0.25 & -1.44 &  -0.58 \\
    SVD & $\boldsymbol{h} \leftarrow \boldsymbol{U}(\boldsymbol{S}+\Delta \boldsymbol{S})\boldsymbol{V}^T\boldsymbol{h} + \mathbf{b}$ &  $\Delta \boldsymbol{S}$  & 192 & 
    -3.05 & -2.48 & -2.52 & -2.68 \\
    adapter  & $\boldsymbol{h} \leftarrow \mathbf{W}\boldsymbol{h}+\mathbf{b} + \mathbf{B}\mathbf{A}(\mathbf{W}\boldsymbol{h}+\mathbf{b})$ & $\mathbf{B}\mathbf{A}$  & 768 & 
    -4.64 & -3.99 & -4.09 & -4.24 \\
    \midrule
    \textbf{low-rank}  & $\boldsymbol{h} \leftarrow (\mathbf{W}+\mathbf{B}\mathbf{A})\boldsymbol{h} + \mathbf{b}$ & $ \mathbf{B}\mathbf{A}$ & 768 & 
    -\textbf{6.38} & -\textbf{5.97} & -\textbf{7.55} & -\textbf{6.63} \\
  \bottomrule 
\end{tabular}
\end{table*} 




{\bfseries Dynamic v.s. Fixed Low-rank Adaptation.}
To verify the effectiveness of our dynamic gate networks, we manually control the number of model update layers and compare that with our dynamic adaptive update. In the experiment, we use the baseline model without model updating as the anchor for computing BD-rate, then we update 1, 2, 4, and 8 layers while our dynamic networks update non-deterministic layers with different image sizes. 

Experiment results are shown in Table \ref{tab:gate1} and reveal that our dynamic model update outperforms pre-specified update layers manually and achieves optimal performance. We can observe that the optimal number of update layers is limited by the input sizes. Though the larger number of update layers brings more improvement in reconstruction quality, the overall RD loss saturates at about four layers of update. Meanwhile, as the image size shrinks, updating fewer weights brings more performance gains, as the additional model streams take up an increasingly large percentage or even over the content streams. It indicates that when the image size is smaller than a certain threshold, the optimal update decision is not to update for the extra rate overhead from model updates does not provide unequal performance gain.



\begin{table}
  \caption{Compare different model update layers with BD-rate ($\%$) on vector arts. The best results are highlighted in bold, while the second-best ones are denoted by underlines.}
  \label{tab:gate1}
  \begin{tabular}{cccc}
    \toprule
    $\#$ of layers & h=256, w=256 & h=384, w=384 & original size \\
    \midrule
    0 & 0.0 & 0.0 & 0.0 \\
    1 & -3.51 & -4.51 & -5.36 \\
    2 & \underline{-3.87}  & -5.90 & -6.92 \\
    4 & -2.77 & \underline{-6.60} & \underline{-8.26} \\
    8 & -1.28 & -5.39 & -7.51 \\
    dynamic & -$\boldsymbol{6.22}$ & -$\boldsymbol{7.84}$ & -$\boldsymbol{8.93}$ \\
  \bottomrule
\end{tabular}
\end{table}
{\bfseries Dynamic Gating Mechanism.} 
To illustrate the intrinsic dynamic gating mechanism, we conduct a statistical analysis of \textit{the frequency of gate opening} of each layer for the model update. We perform our analysis based on Cheng \textit{et al.} \cite{cheng2020learned}, which uses residual blocks and sub-pixel convolution in the decoder to reconstruct the image.
We assign a dynamic network to each residual block for updating of the pre-trained weight and sort them according to the used order. A total of 11 dynamic networks are inserted. 

\begin{figure}
\centering
\includegraphics[width=0.8\linewidth]{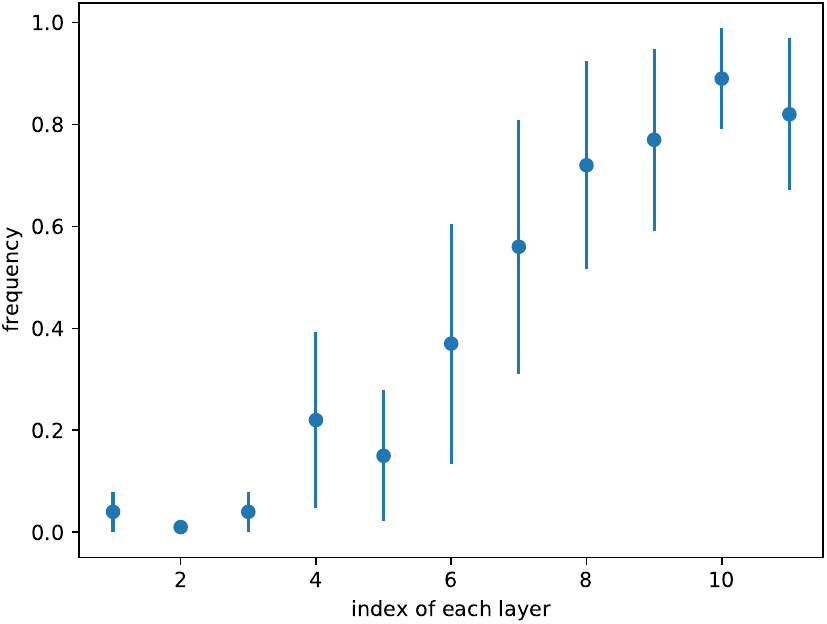} 
\caption{The frequency of each convolution layer is selected for model updates. The index indicates the used order.}
\label{fig:frequency}
\end{figure}

The experimental results, as shown in Fig. \ref{fig:frequency}, indicate that blocks with larger numbers (closer to the model output) are updated more frequently, which is consistent with prior experience. The back convolution layers of the decoder have a small receptive field, making them effective for perceptual detail learning, while the top hidden layers have a larger field and are better suited for learning abstract and semantic features.
The dynamic networks tend to update the layers closer to the reconstructed images due to their ability to construct low-level and pixel information, which is essential for in-domain image reconstruction. 


{\bfseries Application  to  Other  Network  Architectures.}
As depicted in Fig. \ref{fig:deterior}, neural codecs trained on natural images exhibit imperfections in terms of generalization. The gap between training and inference is a common issue encountered by neural codecs. To address this challenge, our dynamic low-rank model adaptation framework is universally applicable to various network architectures. It can effectively reduce the domain gap between training and inference, thereby enhancing the performance of neural codecs.

\begin{figure}
\centering
\includegraphics[width=0.8\linewidth]{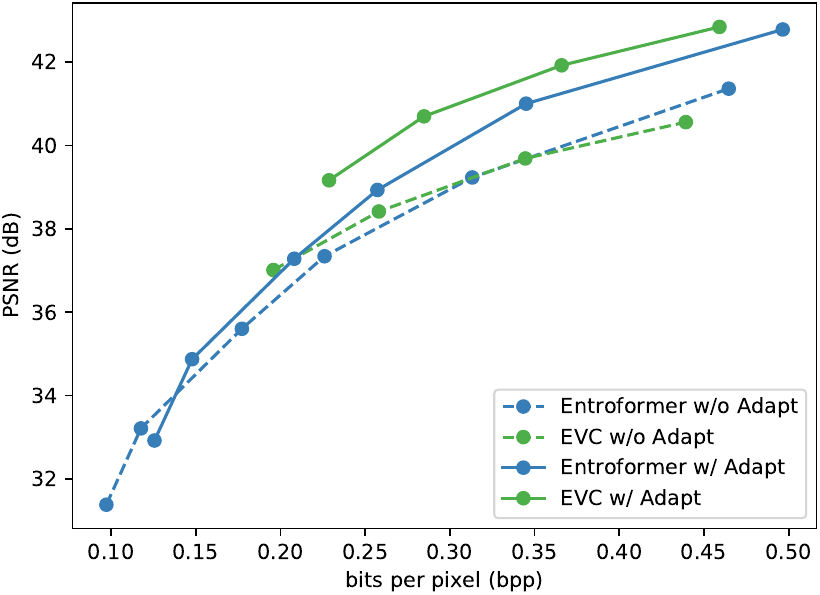} 
\caption{Dynamic adaptation on vector arts based on Entroformer \cite{qian2022entroformer} and EVC\cite{wang2023evc}.}
\label{fig:other}
\end{figure}

We demonstrate the proposed dynamic adaptation without latent refinement when applied to Entroformer \cite{qian2022entroformer} and EVC \cite{wang2023evc}. We adapt the model based on their open-source pre-trained weight. Experiment results are shown in Fig. \ref{fig:other}. 
We have confirmed the universality of our approach and demonstrated the existence of a domain gap in many learning-based compression methods. 
Empirical evidence indicates that the problem under investigation is not the result of individual model defects, but rather a common domain gap between training and inference. 
Our dynamic network addresses this issue by minimizing the domain gap during inference, through minor updates to the deployed neural codecs on the client, rather than the laborious process of fine-tuning all weights for each domain and deploying multiple decompression codecs.

\subsection{Adaptation Efficiency}

Image adaptation techniques, including the one proposed in this study, require additional encoding time and hardware resources to adapt the model. However, this may pose practical limitations in scenarios where compression speed is crucial and computational resources are scarce. To expand the utility, we propose a one-step adaptation approach with one forward during inference. Please refer to the \textbf{Supplementary} for the one-step adaptation setup.

Table \ref{tab:time} shows the correlation between the adaptation performance and the inference time using our method. Notably, adaptive image compression techniques yield superior rate-distortion (RD) performance, albeit at the expense of increased inference time. Our evaluation of inference time was conducted on a single NVIDIA Tesla V100 GPU. 

\begin{table}
  \caption{ Correlation between the adaptation performance and the inference time on vector art images.}
  \label{tab:time}
  \begin{tabular}{cccccccc}
    \toprule
    Steps & 1 & 100 & 500 & 1000 & 2000 & 3000 & 4000 \\
    \midrule
    Time & 0s & 13s & 55s & 110s & 219s & 322s & 427s \\
    \midrule
    BD-rate & -1.07 & -3.77 & -7.47 & -9.43 & -11.14 & -17.17 & -17.49 \\
     
  \bottomrule
\end{tabular}
\end{table}
\section{CONCLUSION}

This study presents a dynamic low-rank instance-adaptive image compression framework that aims to mitigate the domain gap and prevent performance degradation on out-of-domain images. The proposed approach involves incremental updates to the pre-trained weight through the use of low-rank decomposition matrices to minimize the rate overhead. Moreover, a dynamic network is designed to enable adaptive control over the number and position of plug-in blocks. However, a major limitation of this study is the additional encoding time required for adaptation. It should be noted that the fast one-step adaptation falls short of achieving full adaptation. These findings offer valuable insights for future research on real-time inference for universal instance adaptive neural image compression.



\begin{acks}
This work was partly supported by the National Natural Science Foundation of China (Nos. 62171251\& 62311530100), the Special Foundations for the Development of Strategic Emerging Industries of Shenzhen (Nos. JCYJ20200109143010272 \& CJGJZD20210408092804011) and Oversea Cooperation Foundation of Tsinghua.
\end{acks}

\newpage
\bibliographystyle{ACM-Reference-Format}
\bibliography{sample-base}

\appendix
\section{Supplementary Material}

\subsection{One-Step Adaptation}

Instance adaptive image compression techniques require additional encoding time to adapt the model, resulting in better rate-distortion (RD) performance. Nonetheless, this approach may not be suitable for certain practical usage scenarios, especially those that prioritize compression speed. To broaden the scope of instance adaptive image compression applications, this section introduces a one-step method that eliminates the need for iterative overfitting of instance parameters during inference, while still slightly enhancing compression performance.

Directly obtaining instance parameters without iteration is not feasible, but we can obtain an approximation by analyzing similar images. To achieve this, we collected 7,000 images from various sources, including the BAM dataset \cite{wilber2017bam}, which were not previously used in evaluating the method. 

After that, we use open-source pre-trained HCSC \cite{guo2022hcsc} to extract features: we take the final layer outputs of a pre-trained HCSC and embed these images into 2048-dimensional feature space. Then we unsupervised clustered these embedding features with $N_{clusters} = 128$ and save the cluster center $c_n,  n \in \{1, 2, \cdots, N_{clusters}\}$. Images in the same clusters usually have similar image textures. For each cluster $\mathcal{C}_n$, we fine-tune incremental matrices $\boldsymbol{\Delta \theta}^n$ on the penultimate layer in the decoder. 

At inference time, for each input $\boldsymbol{x}$, we extract its embedding feature $\mathcal{F}(\boldsymbol{x})$ and find the nearest cluster center $n$: 
\begin{equation}
    n = \arg \min d(\mathcal{F}(\boldsymbol{x}), c_n)
\end{equation}

The corresponding parameters $\boldsymbol{\Delta \theta}^n$ of $\mathcal{C}_n$ are quantized and decoded to bit rate and then transmitted to the client to enhance the reconstruction equality:
\begin{equation}
    \hat{\boldsymbol{x}} = g_s(\boldsymbol{\hat y}; \boldsymbol{\theta}, \boldsymbol{\Delta \hat \theta }^n)
\end{equation}

 Our evaluation of one-step adaptation includes in-domain and out-of-domain images using the Cheng \textit{et al.} \cite{cheng2020learned} baseline as an anchor for computing BD-rate. To compare, we incrementally update the penultimate layer in the decoder for each cluster. The experiment results, presented in Table \ref{tab:free}, demonstrate that performance can be improved even without adaptation during inference time, on both in-domain and out-of-domain images. In strict terms, one step is a complete process containing end-to-end forward propagation and backward propagation. Our one-step adaptation just forwards the encoding process and then matches the adaptation parameters. One-step adaptation reduces the inference time and server performance requirements, making these methods more practical.
 

\begin{table}[h]
  \caption{One-step adaptation  results in BD-rate (\%).}
  \label{tab:free}
  \begin{tabular}{ccccc}
    \toprule
    Method & Kodak & Vector Art & Comic & Pixel Art \\
    \midrule
    Cheng \textit{et al.} & 0.0 & 0.0 & 0.0 & 0.0   \\
    Cheng \textit{et al.} + Ours & -0.24 & -1.07 & -1.79 & -0.50   \\
  \bottomrule
\end{tabular}
\end{table}

\subsection{Extend to Pathology Images}
Digital pathology images are limited in their application in areas such as remote collaborative diagnosis and AI-assisted diagnosis due to their ultra-high resolution and expensive save cost. Meanwhile, despite the neural codecs developing rapidly, most of them are trained on natural images, and there is no neural codec specifically for pathology images. We extend this study to pathology images in an attempt to fill this blank. 

We evaluate our method on pathology images from BRACS \cite{brancati2022bracs} dataset. 
We compare our method with open-source instance-adaptive image compression methods. The evaluation measured in terms of BD-rate (\%) using VVC as the anchor. The experiment results are presented in Table \ref{tab:pathology}. 

\begin{figure}[h]
\centering
\includegraphics[width=0.8\linewidth]{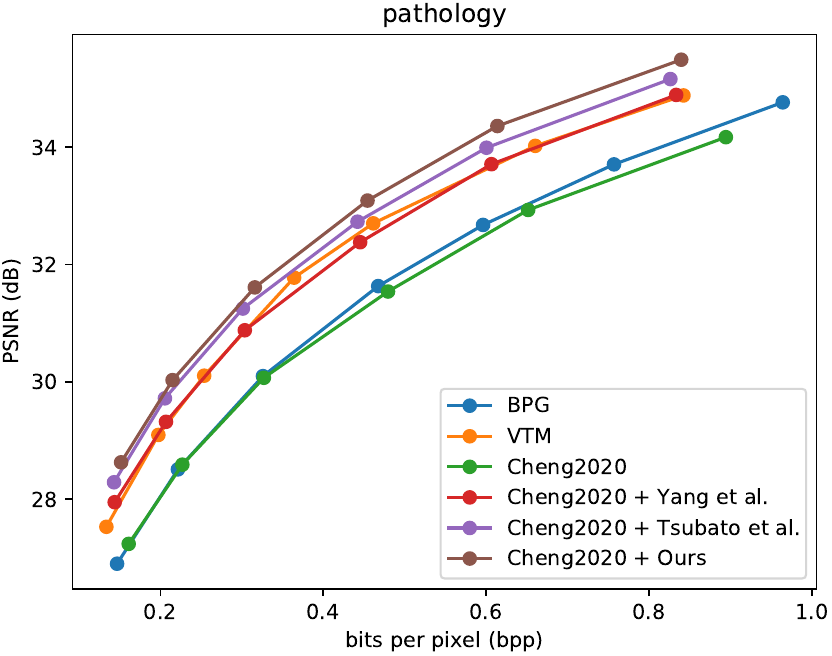} 
\caption{The RD performance with BPG, VVC, and existing instance-adaptive image compression methods on pathology images.}
\label{fig:pathology}
\vspace{-1em}
\end{figure}

\begin{table}[h]
  \caption{Comparison on pathology images with BPG, VVC and other existing adaptive approaches. The evaluation is measured in terms of BD-rate (\%) using VVC as the anchor.}
  \label{tab:pathology}
  \begin{tabular}{cccccc}
    \toprule
    Method& BD-rate (\%) \\
    \midrule
    BPG                     
    &  10.56 \\
    VVC                     
    & 0.00  \\
    Cheng2020 \cite{cheng2020learned}            
    & 11.63 \\
    Cheng2020 + Yang \textit{et al.} \cite{yang2020improving}            
    & 0.42 \\
    Cheng2020 + Tsubota \textit{et al.}  \cite{tsubota2023universal}        
    &  -3.02 \\
    Cheng2020 + Ours                    
    & -$\boldsymbol{5.04}$ \\
  \bottomrule
\end{tabular}
\end{table}

During inference time adaptation, tiny updates to the deployed neural codecs on the client greatly compensates the domain gap. Our instance adaptive image compression model avoids preparing and deploying multiple task-specific decompression codecs. Images from any domain can be compressed by only one base compression model while alleviating the performance degradation and achieving outstanding performance.

\subsection{Visualization of the Model Update}
\begin{figure*}
\centering
\includegraphics[width=180mm]{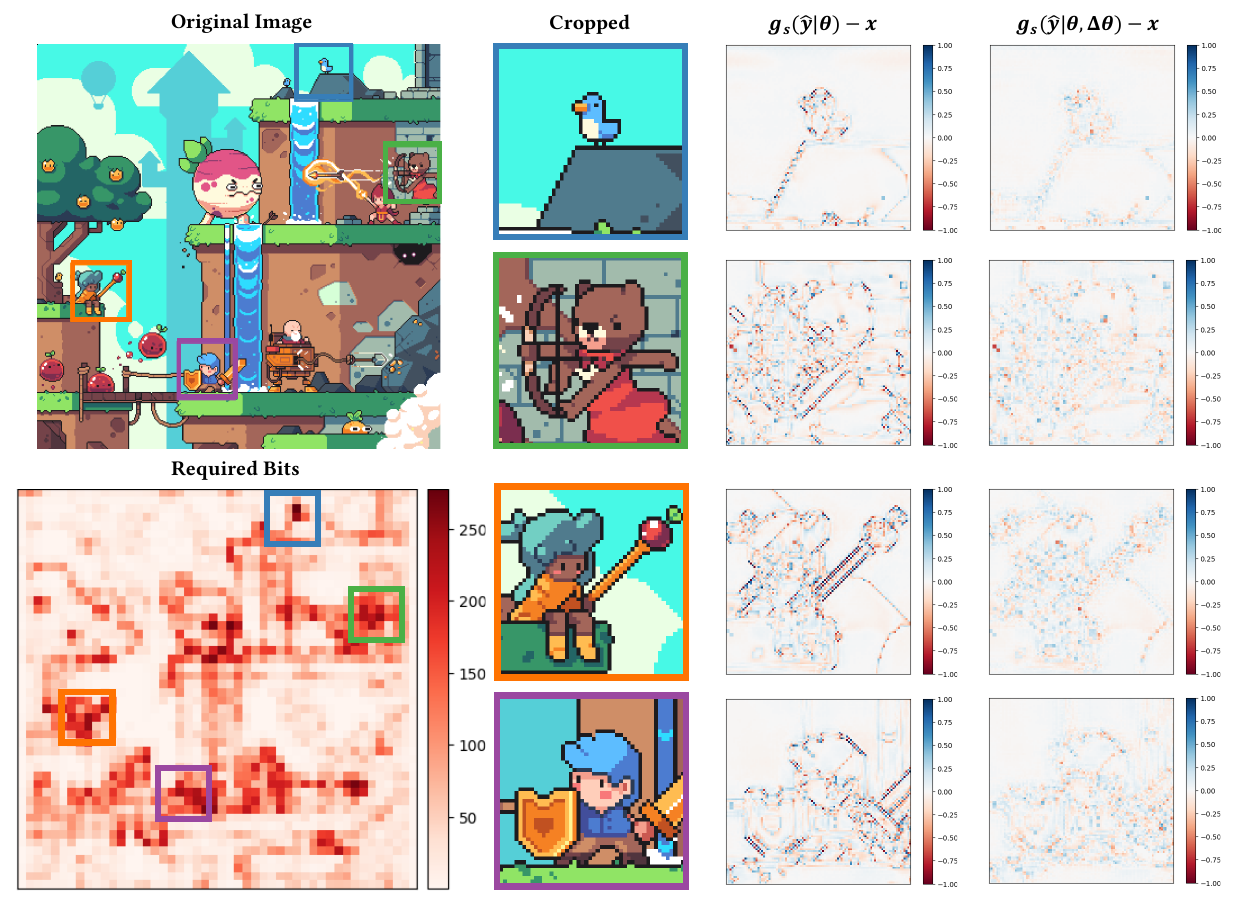}
\caption{Visualization of reconstruction with and without extra parameters $\boldsymbol{\Delta \hat \theta}$ using a pixel-style image as an example. It shows our approach provides domain-specific image texture to reduce reconstruction error significantly.}
\label{fig:vis}
\end{figure*}

We reconstruct the refined quantized latent code with and without extra parameters $\Delta \boldsymbol{\hat \theta}$ to demonstrate its effectiveness. Required bits are calculated as $ -\log_2 p(\boldsymbol{\hat y})$ to show the bit allocation. We choose four patches that require more bits in the original images and then reconstruct the latent code without the model update as $\boldsymbol{\hat x} =g_s(\boldsymbol{\hat y}; \boldsymbol{\theta})$ and with extra parameters $\Delta \boldsymbol{\hat \theta}$  as $\boldsymbol{\hat x} =g_s(\boldsymbol{\hat y}; \boldsymbol{\theta}, \boldsymbol{\Delta} \boldsymbol{\hat \theta}$), respectively. Then we visualize the reconstruction error $\boldsymbol{\hat x} - \boldsymbol{x}$. As shown in Fig. \ref{fig:vis}, we can observe that the error distributes mainly at edges and our approach reduced the reconstruction errors significantly and provides domain-specific image texture during adaptation.

\subsection{Visualization Comparison with Other Approaches} 
\begin{figure*}
\centering
\includegraphics[width=1\linewidth]{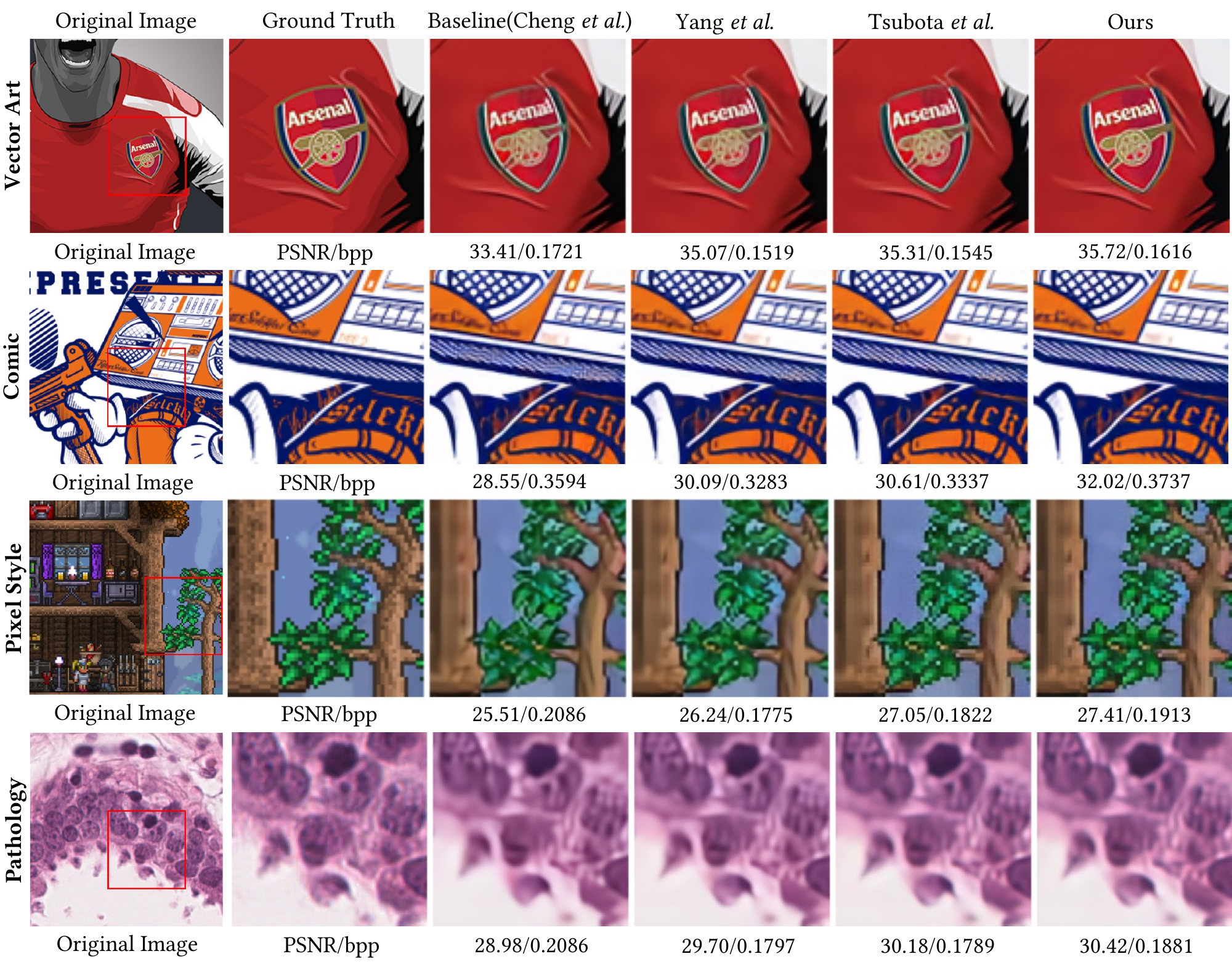}
\caption{Visualization and PSNR/bpp performance comparison with existing instance-adaptive image compression methods Yang \textit{et al.} \cite{yang2020improving} and Tsubota \textit{et al.} \cite{tsubota2023universal} based on the same pre-trained baseline cheng \textit{et al.}\cite{cheng2020learned} from compressai \cite{begaint2020compressai}. The PSNR/bpp metrics are under each patch. Zoom in for better visibility.}
\label{fig:vis2}
\end{figure*}

The qualitative results of the baseline method Cheng \textit{et al.} \cite{cheng2020learned} without adaptation, other instance-adaptive image compression methods \cite{yang2020improving,tsubota2023universal}, and ours are shown in Fig. \ref{fig:vis2}. 
We conducted a comparative analysis of these techniques utilizing the same pre-trained model on three distinct out-of-domain image categories: vector art, comic, and pixel style. Our approach demonstrated superior visual quality in comparison to both the baseline method and other instance-adaptive codecs.

Our approach in the vector art patch yields superior visual quality when reconstructing patterns and text, resulting in clearer boundaries and higher fidelity. In the case of the comic example, the baseline method without adaptation struggles with reconstructing dense diagonal lines. While other adaptation methods reduce the reconstruction error, they still exhibit some artifacts. Our method successfully reconstructs dense lines with adaptation. We used a pixel-style image from Terraria as a sample, and the reconstruction results demonstrate that the baseline method without adaptation results in smoothing and chromatic aberration. Furthermore, it fails to reconstruct jagged edges and instead produces smooth edges, which do not accurately capture the characteristics of pixel-style images. However, our method, with model adaptation, results in better restoration of color and pixel-level edges. We randomly select a breast histopathological image from BRACS \cite{brancati2022bracs} as a sample to show that our method retain more nuclei details than the method without adaptation.

\subsection{Generalization on Invertible Neural Networks}
\begin{figure*}
\centering
\includegraphics[width=1\linewidth]{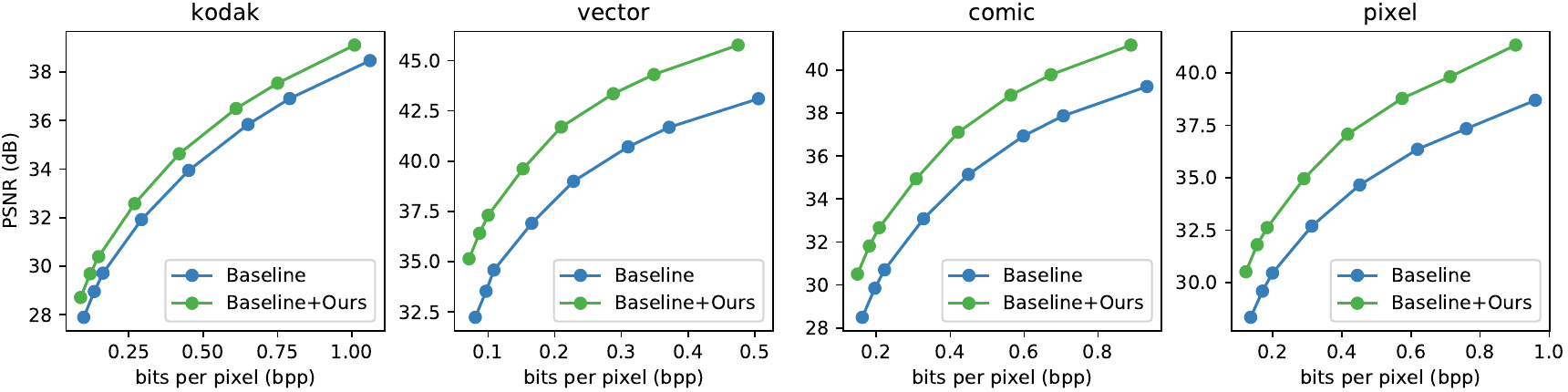}
\caption{The RD performance of INN-based image compression methods with and without instance adaptation. BD rates are shown in Table \ref{tab:inn}.}
\label{fig:inn}
\end{figure*}

To show the generalization potential of our dynamic instance image compression method, we implement instance adaptive compression based on invertible neural network (INN) \cite{xie2021enhanced} backbone. Reconstruct network is updated by low-rank decomposition matrices to enhance compression performance. The evaluation measured in terms of BD-rate (\%) using baseline as the anchor is presented in Table \ref{tab:inn} and Fig. \ref{fig:inn}

\begin{table}[h]
  \caption{Evaluation results based on INN-based method \cite{xie2021enhanced}.}
  \label{tab:inn}
  \begin{tabular}{ccccc}
    \toprule
    Method & Kodak &Vector art&Comic& Pixel art \\
    \midrule
    INN & 0.0 & 0.0 & 0.0 & 0.0 \\
    INN + Ours & -8.42 & -10.43 & -13.94 & -18.08 \\
  \bottomrule
\end{tabular}
\end{table}

\end{document}